\documentclass[aps,prl,twocolumn,superscriptaddress,showpacs,floatfix]{revtex4}
\usepackage{graphicx}
\usepackage{color}

\usepackage{amssymb}

\begin{document}

\title{Experimental test of quantum non-Gaussianity of heralded single photon state}

\author{Miroslav Je\v{z}ek}
\author{Ivo Straka}
\author{Michal Mi{\v{c}}uda}
\author{Miloslav Du{\v{s}}ek}
\author{Jarom\'{\i}r Fiur\'{a}\v{s}ek}
\author{Radim Filip}

\affiliation{Department of Optics, Palack\'y University, 17.~listopadu 12, 77146 Olomouc, Czech Republic}

\begin{abstract}
We report on experimental verification of quantum non-Gaussianity
of a heralded single photon state with positive Wigner function. We unambiguously demonstrate that the
generated state cannot be expressed as a mixture of Gaussian states.
A sufficient information to witness the quantum non-Gaussianity is
obtained from a standard photon anti-correlation measurement.
\end{abstract}


\maketitle

Quantum properties of light are exemplified by statistical behaviors
which do not admit explanation based on a semiclassical theory.
Since coherent states represent quantum analogue of  classical coherent light, a state that cannot be expressed as
 a convex mixture of coherent states is commonly considered to be nonclassical \cite{Glauber}.
In particular, during recent decades  nonclassical squeezed states of light have 
become a crucial resource for quantum optics, metrology, and quantum information processing \cite{Furusawa,Bachor}.

Pure squeezed coherent states represent extremal points of a convex set of stochastic mixtures of Gaussian
states. All such states possess positive Wigner function and can be obtained from coherent laser beams using
classical mixing and quantum interactions described by quadratic Hamiltonians. Exploiting higher-order nonlinearities involved in the photon detection process,
states with negative Wigner function can be conditionally generated from the squeezed states \cite{Lvovsky01,Bellini04,Ourjoumtsev06,Polzik06,Sasaki10}.
Wigner function of these highly non-classical states exhibits a distinctly non-Gaussian shape
that cannot be obtained as a stochastic mixture of Gaussian functions.

The famous Hudson theorem establishes an equivalence between the non-Gaussianity and negativity of Wigner function for pure states \cite{Hudson74}.
However, this relation does not simply extend to mixed states \cite{Cerf09}. Previous approaches towards the non-Gaussianity witness or measure for mixed states
do not distinguish non-Gaussianity which is compatible with a simple mixture of Gaussian states, and they also require complete 
information about the quantum state \cite{Dodonov00,Genoni08,Barbieri10}.
This  brings a very basic and fundamental physical problem to our attention:
Which mixed non-classical quantum states with positive non-Gaussian Wigner function
do not admit explanation based solely on stochastic non-Gaussianity? 
Mathematically, we search for a witness certifying that the state cannot be constructed as a mixture of Gaussian states.
Very recently, a directly measurable witness of the quantum non-Gaussianity
has been theoretically proposed \cite{Filip11}.
The witness is based on knowledge of probabilities of vacuum and single-photon states only, yet it can detect
a wide class of states with positive Wigner function which are not mixtures of Gaussian states.

A heralded single photon source is an excellent example for testing the power of this witness in a laboratory.
In the absence of background noise, the generated state would be a mixture of a single photon state and a vacuum due to losses and imperfect coupling
and mode-matching.
If the probability of vacuum dominates, then the state exhibits positive Wigner function. Nevertheless, the witness \cite{Filip11} still proves
that it is not a mixture of Gaussian states. Here we apply the witness to approximate single-photon states conditionally
generated by detection of an idler photon from a photon pair produced by the process of spontaneous parametric frequency down-conversion (PDC).
Our detection scheme consisting of a beam splitter and two single-photon detectors is the one commonly employed to test the anticorrelation properties 
of single-photon sources \cite{Grangier86}. This measurement allows us to obtain suitable estimates of vacuum and single-photon probabilities, 
which are required for the non-Gaussianity witness. The verification of quantum non-Gaussianity 
thus conveniently complements other typically performed non-classicality tests of single photon sources.

{\em Theory:} Let $\mathcal{G}$ denote the set of all mixtures of Gaussian states.
We would like to show that a given state $\rho$ cannot be expressed as a convex mixture of Gaussian states, $\rho \notin \mathcal{G}$,
even though $\rho$ possesses a positive Wigner function. This can be accomplished using a criterion recently derived in Ref. \cite{Filip11}.
This criterion is based on photon number probabilities and can be expressed as an upper bound on single-photon probability $p_1$
for a given vacuum state probability $p_0$. If the measured $p_1$ exceeds this bound, then $\rho \notin \mathcal{G}$.
The bound can be derived by maximizing $p_1$ for a fixed $p_0$ over all pure Gaussian states  \cite{Filip11} and can be
conveniently expressed in a parametric form,
\begin{eqnarray}
p_0=\frac{e^{-d^2[1-\tanh(r)]}}{\cosh(r)}, \qquad
p_1=\frac{d^2 \, e^{-d^2[1-\tanh(r)]}}{\cosh^3(r)}.
\label{p01bound}
\end{eqnarray}
Here $r \geq 0$ is the squeezing constant and the displacement reads $d^2=(e^{4r}-1)/4$.
All probability pairs ($p_0,p_1$) achievable by mixtures of Gaussian states form a convex set that is shown in Fig. 1(a) as a blue area.
Note that the boundary of this area is specified by the formula (\ref{p01bound}).

In analogy with entanglement witnesses \cite{Lewenstein00}, we can define a non-Gaussianity witness \cite{Filip11},
\begin{equation}
W(a)=ap_0+p_1.
\label{Wa}
\end{equation}
If  $W(a)> W_G(a)$ then $\rho \notin \mathcal{G}$. The bound $W_{G}(a)=\max_{\rho\in\mathcal{G}}W(a)$ can be obtained
by solving the equation $(1+e^{2r})a=e^{2r}(3-e^{2r})$ with respect to $r$,
determining $p_0$ and $p_1$ from Eq. (\ref{p01bound}) and inserting them into
Eq. (\ref{Wa}). As indicated by a green dashed line in Fig.~1(a), each line $ap_0+p_1=W_G(a)$ is a tangent to the boundary curve (\ref{p01bound})
and divides the plane into two half-planes. All points $(p_0,p_1)$
lying in the half-plane $ap_0+p_1>W_G(a)$ are certified by the witness to correspond to a state $\rho \notin \mathcal{G}$.

\begin{figure}[b]
\includegraphics[width=0.95\linewidth]{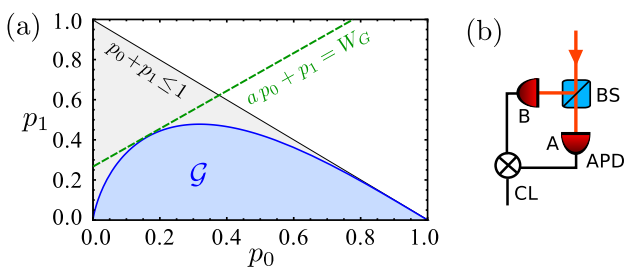}
\caption{(Color online) (a) Inferring non-Gaussianity from photon number probabilities $p_0$ and $p_1$
of vacuum and single-photon states. The region of physically allowed points $(p_0,p_1)$
is formed by a triangle $p_0+p_1\leq 1$, $p_j\geq 0$. The blue convex region ${\cal G}$
represents probability pairs achievable by mixtures of Gaussian states. The points lying
in the light gray region indicate states that cannot be expressed as convex mixtures of Gaussian
states. The green dashed line represents one of the non-Gaussianity witnesses $W(a)$.
(b) Detection scheme. Signal light beam impinges on a beam splitter (BS) and the outputs are detected by two
APDs. Both single and coincidence rates are acquired by coincidence logic (CL).}
\end{figure}

Let us now consider practical determination of the probabilities $p_0$ and $p_1$. 
Since the currently commonly available avalanche photodiodes (APDs) are not capable of resolving the number of photons,
one needs to employ an advanced photon-number resolving detector \cite{Kim99,Lita08,Kardyna08} 
with demanding operation conditions or some sort of multiplexed detector \cite{Paul96,Banaszek00,Rehacek03,Achilles03,Fitch03,Micuda08,Avenhaus10,Kalashnikov11}. 
Perhaps the conceptually simplest scheme,
shown in Fig. 1(b), is based on splitting the incoming signal on a balanced beam splitter 
BS and placing an APD on each output port of the BS. This setup is commonly used for measurement of the $g^{(2)}$ factor \cite{Grangier86,Weihs09,Benson11}.

Although this scheme provides some information about photon statistics there are several factors that need to be carefully considered.
One important issue is the detector efficiency and other losses which combine to overall efficiency $\eta$. Compensation of $\eta$ would require its
precise calibration, which is a non-trivial task. However, we can simply include losses into state preparation. Let $L_{\eta}$
denote a lossy channel with transmittance $\eta$.
This channel maps Gaussian states onto Gaussian states. Therefore, if $\rho \in \mathcal{G}$ then also
$\rho_\eta \equiv L_{\eta}[\rho] \in \mathcal{G}$. This implies that if $\rho_\eta \notin \mathcal{G}$ then also $\rho \notin \mathcal{G}$. 
We can thus conservatively
assume perfect detectors with unit efficiency and if $\rho_\eta \notin \mathcal{G}$ is proven under this assumption, then it certainly holds also 
for $\rho$ irrespective of the exact value of $\eta$.

In the experiment, the number of single detector clicks ($R_{1A}$ and $R_{1B}$) as well as number of the coincidence clicks ($R_2$)
is measured for a given number of samples $R_0$
of the state. Assuming perfect detectors with  $\eta=1$ the vacuum-state fraction $p_0$ is the probability that none of the detectors clicks,
\begin{equation}
p_0=1-\frac{R_{1A}+R_{1B}+R_2}{R_0}.
\label{p0est}
\end{equation}
The determination of $p_1$ is less trivial. We have
\begin{equation}
\frac{R_{1A}}{R_0}=\sum_{n=1}^\infty T^n  p_n, \qquad \frac{R_{1B}}{R_0}=\sum_{n=1}^\infty (1-T)^n p_n,
\label{Rfraction}
\end{equation}
where $T$ denotes the effective transmittance of the BS that also includes possible imbalance of the detection efficiencies and other factors.
Note that $R_{1A}$ and $R_{1B}$ depend on  the whole photon number distribution $p_n$, not just on $p_0$ and $p_1$. We can nevertheless construct
the following estimator,
\begin{equation}
p_{1,\mathrm{est}}=\frac{R_{1A}+R_{1B}}{R_{0}}- \frac{T^2+(1-T)^2}{2T(1-T)}\frac{R_{2}}{R_0}.
\label{p1est}
\end{equation}
With the help of Eqs. (\ref{p0est}) and (\ref{Rfraction}) one can show that
\[
p_{1,\mathrm{est}}=p_1- \sum_{n=3}^\infty p_n\frac{T^2-T^n+(1-T)^2-(1-T)^n}{2T(1-T)},
\]
hence $p_{1,\mathrm{est}} \leq p_1$. Note that the term proportional to $p_2$ is absent in $p_{1,\mathrm{est}}$,
so for rapidly decaying distributions the error is of the order of $p_3$.
With this lower bound on $p_1$ at hand, the above criterion is still
applicable, because $ap_0+p_{1,\mathrm{est}}>W_G(a)$ implies that $ap_0+p_1>W_G(a)$ as well.

The estimation of $p_1$ is influenced by the effective imbalance of the detection channels $T:(1-T)$. Without loss of generality we can assume that $T>\frac{1}{2}$.
It follows from Eq. (\ref{p1est}) that $p_{1,\mathrm{est}}$ decreases with increasing ratio $T:(1-T)$. Hence we should avoid underestimation of $T$ which
would result in overestimation of $p_{1}$. An upper bound on $T$ is provided by  the ratio of single detector clicks,
\begin{equation} \label{spliratioest}
T_{\mathrm{est}}=\frac{R_{1A}}{R_{1A}+R_{1B}}.
\end{equation}
It can be shown that  $ T \leq T_{\mathrm{est}}$ for $T \geq \frac{1}{2}$. We can thus safely use $T_{\mathrm{est}}$
as a conservative estimate of $T$.

{\em Multimode witness:}
Many single-photon sources do not emit photons strictly into a single spatial and temporal mode. Let us therefore briefly sketch a proof \cite{Filip_multimode} 
of the applicability of the witness to a generic multi-mode case. Let $N$ denote the total number of modes involved and we define the
total photon number $n= \sum_{j=1}^N n_j$, where $n_j$ is the number of photons in $j$-th mode.
The estimated probabilities $p_n$ then correspond to probability of no photon ($n=0$) or one photon in total ($n=1$) in the signal beam.
Even in this multimode case, the maximum of $W(a)=ap_0+p_1$ over all mixtures of Gaussian states is attained by a pure
$N$-mode Gaussian state. 
Any $N$-mode pure Gaussian state can be prepared by combining $N$ single-mode squeezed states in a network of beam splitters. Moreover, the passive linear
network described by a unitary matrix $U_N$ does not change the statistics of total photon number because $U_N n U_N^\dagger =n$.
It therefore suffices to carry out the optimization over products of $N$ pure single-mode Gaussian states which can be done analytically and 
one can prove that the bound on $W(a)$ remains $W_G(a)$ for arbitrary $N$. We can therefore apply the witness also
to multi-mode states without any limitation.

\begin{figure}[!b]
\centerline{\includegraphics[width=0.9\columnwidth]{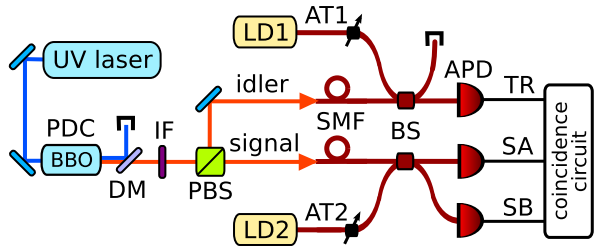}}
\caption{(Color online) Layout of the experimental setup.
Continuous-wave frequency-multimode ultraviolet laser with central
wavelength of $407$~nm pumps a $2$~mm long $\beta$-barium borate (BBO)
nonlinear crystal phase matched for type-2 degenerate PDC. Collinearly generated photon pairs
with central wavelength of $814$~nm are collimated and separated
from the pump beam by a dichroic mirror (DM) and spectrally limited
by an interference filter (IF) to $10$~nm.
The orthogonally polarized photons of a PDC pair are separated
by a polarizing beam splitter (PBS) and coupled to single-mode
optical fibers (SMF). Both outputs can be mixed with an attenuated (AT)
infrared laser diode (LD) signal at fiber beam splitters (BS) to emulate dark
counts of detectors and a noise component of the explored quantum state.
The three output modes are then detected by binary single photon detectors
based on silicon avalanche photodiodes (APD) operated in Geiger mode and actively quenched.
The absolute quantum efficiency of the detectors is specified by
a manufacturer to approximately $50\%$, while their relative
efficiencies were precisely measured prior to the experiment and
found to be $100\%,\,91.7\pm0.2\%,\,92.2\pm0.2\%$ for
TR,\,SA,\,SB channels, respectively. Electronic dark counts
of the detectors in coincidence basis were found to be completely
negligible.}
	\label{fig:exp_setup}
\end{figure}

{\em Experimental setup:}
The experimental setup of a heralded single photon source based on PDC
is presented in Fig.~\ref{fig:exp_setup}. The detection part consists
of three binary detectors (TR,\,SA,\,SB).
The trigger detector (TR) yields a heralding output of the
single photon PDC source. When a detection event is registered
at this detector, we expect that an approximate single photon state
is prepared in signal mode. The signal is divided by the beam
splitter (BS) with transmittance of $T_{BS}=0.522\pm0.003$ to the detection
channels SA and SB. All single as well as two-fold and three-fold
coincidence events between channels TR, SA, and SB are registered
by a fast coincidence logic unit. The overall splitting ratio $T_{\mathrm{est}}$ between the channels SA and SB has been conservatively estimated
from the measured rates $R_{1A}$ and $R_{1B}$ using Eq. (\ref{spliratioest}) and it agrees well with
the independently measured $T_{BS}$ and relative detector efficiencies.

\begin{table}[t]
\caption{Estimated probabilities $p_0$ and $p_1$, and the corresponding witness $\Delta W$ are shown for several different pump powers $P$ and IF widths
$w$ ($-$ denotes no filter).}
\begin{ruledtabular}
\begin{tabular}{ccrr@{}r@{\quad}}
\multicolumn{1}{c}{$P$\,[mW]}& \multicolumn{1}{c}{$w$\,[nm]} &
\multicolumn{1}{c}{$p_0$} & \multicolumn{1}{c}{$p_1$} &
\multicolumn{1}{c}{$\Delta W\,[\times10^{-6}]$}  \\ \hline
50 & 2   & $0.9124$ & $0.0875$ & $412 \pm 1~\,$  \\
50 & 10  & $0.8589$ & $0.1410$ & $1666 \pm 3~\,$  \\
20 & 10  & $0.8425$ & $0.1574$ & $2370 \pm 2~\,$ \\
50 & $-$ & $0.7095$ & $0.2901$ & $14252 \pm 17$  \\
5  & $-$ & $0.7296$ & $0.2704$ & $11825 \pm 15$ 
\end{tabular}
\end{ruledtabular}
\end{table}

The probabilities $p_0$ and $p_1$ are estimated from the measured data using formulas (\ref{p0est}) and (\ref{p1est}).
Due to conditioning on clicks of the trigger, $R_0$ is given by the singles of the trigger detector,
$R_{1A}$ and $R_{1B}$ by two-fold coincidences of TR\&{SA} and TR\&{SB}, respectively, while $R_2$ is actually given by the three-fold coincidences.
The coincidence window is set to $2$~ns. 
The results are summarized in Table I for different pump powers $P$ and three different
full width at half maximum (FWHM) of the interference filter IF ($2$~nm, $10$~nm and without filter).
Due to imperfect mode matching, in-coupling losses and inefficient detectors, the vacuum term $p_0$ dominates while the single-photon fraction is below $30\%$ for all data shown.
The contribution of higher photon terms is very small, $1-p_0-p_{1} \lesssim 10^{-4}$, so the generated state is very close to an attenuated single photon.
In the experiment, this is indicated by a very small ratio of three-fold to two-fold coincidence rates (less than $10^{-3}$).
For example, for $P=50$~mW and $w=10$~nm we have $R_2=605$, $R_{1A}=1.259\times10^6$ and $R_{1B}=1.192\times10^6$ per $100$~s.
The statistical uncertainty of the estimated $p_0$ and $p_1$ determined assuming Poissonian statistics is less than $2\times 10^{-4}$
for all data shown (one standard deviation). Due to very low three-fold coincidences, the sum $p_0+p_1$ exhibits much lower 
statistical uncertainty than the difference $p_0-p_1$.

{\em Results:}
We have verified that the generated states cannot be expressed as a mixture of Gaussian states by using the non-Gaussianity witness.
For each data set $(p_0,p_1)$
we have calculated $\Delta W=a p_0+p_1 -W_G(a)$ and maximized the difference over all $a$.
The resulting maximal $\Delta W$ are listed in Table I. We can see that $\Delta W >0$ in all cases and the  
bound $W_G(a)$ is always surpassed by many standard deviations. 
Next, we investigate the influence of background noise on the source properties.
For this purpose we inject light from laser diodes LD1 and LD2 into trigger and signal detection blocks, respectively. Noise from LD2
emulates background noise of the source while noise coming from LD1 effectively increases dark count rate of the trigger 
thus increasing the vacuum fraction $p_0$.
Table II  shows the results obtained when the amount of injected noise is the same in both blocks and $n_{\mathrm{rel}}$ indicates the normalized noise strength.
With increasing noise we can clearly observe transition to the regime where $\Delta W<0$, as also illustrated in Fig. 3.

\begin{table}[t]
\caption{The same as Table I, $n_{\mathrm{rel}}$ indicates the amount of noise injected from LD1 and LD2
into trigger and signal detectors, $P=50$ mW, and IF width $w=10$~nm.}
\begin{ruledtabular}
\begin{tabular}{@{~~}ccccr@{\qquad}}
$n_{\mathrm{rel}}$ &$p_0$ & $p_1$ &$a_{\mathrm{opt}}$ &
\multicolumn{1}{c@{\quad}}{$\Delta W\,[\times10^{-6}]$} \\ \hline
 0.0 &        0.8195        & 0.1804 &        0.94018        &  $3479 \pm 7$ \\
 0.1 &        0.9073        & 0.0926 &        0.98389        &   $406 \pm 3$ \\
 0.2 &        0.9408        & 0.0591 &        0.99332        &    $42 \pm 2$ \\
 1.0 &        0.9777        & 0.0222 &        0.99903        &   $-84 \pm 1$ \\
\end{tabular}
\end{ruledtabular}
\end{table}

\begin{figure}[b]
\includegraphics[width=0.8\linewidth]{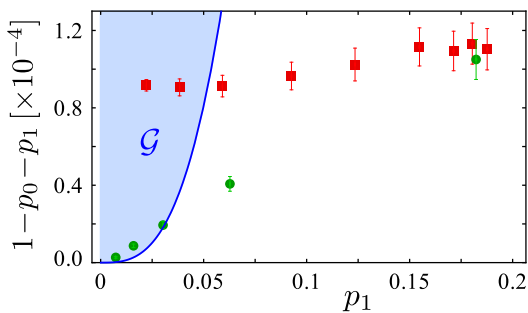}
\caption{(Color online) The multiphoton contribution $1-p_0-p_1$ is plotted as a function of $p_1$.
Symbols indicate experimental data for several different levels of noise added simultaneously by LD1 and LD2 (red squares) and by LD1 only (green circles). 
All other parameters were fixed, $P=50$~mW and $w=10$ nm. Error bars stand for three standard deviations, horizontal error bars of $p_1$ are smaller than the size
of the symbols. The solid blue curve represents the boundary given by Eq. (\ref{p01bound}),
all points on the right of this curve correspond to states that cannot be obtained as mixtures of Gaussian states.}
\end{figure}

{\em Discussion:}
Let us briefly compare our results with other non-classicality measures. Since $p_0>0.5$ for all the measured states,
their Wigner function is always positive in the origin, $W(0)=\frac{1}{\pi}\langle (-1)^{n}\rangle\geq \frac{2p_0-1}{\pi} >0$, where it
is expected to exhibit maximum negativity  $W(0)=-\frac{1}{\pi}$ for a pure single photon state \cite{Silberhorn}.
On the other hand, all the measured states cannot be expressed as a mixture of coherent states, therefore
they are non-classical \cite{note1}.
The non-classicality can be quantified by a $g^{(2)}$ parameter defined as
$g^{(2)}(0)=\frac{\langle a^{\dagger\,2}a^{2}\rangle}{\langle a^{\dagger}a\rangle^2}$ \cite{note2}.
Sub-Poissonian photon number statistics is indicated by  $g^{(2)}(0)<1$. 
The state produced by our source can be excellently approximated
by a density matrix $\rho_T=p_0|0\rangle\langle 0|+p_1|1\rangle\langle 1|+(1-p_0-p_1)|2\rangle\langle 2|$ because higher photon terms
are exponentially suppressed due to very low parametric gain in the nonlinear crystal. In this case we find
$g^{(2)}(0)=\frac{2(1-p_1-p_0)}{[2(1-p_0)-p_1]^2}$ yielding  $g^{(2)}(0)<0.3661$  for all states.
Simultaneously,  all the results exhibit very
strong photon anti-correlation effect, witnessed by $\alpha=\frac{R_0R_2}{R_{1A}R_{1B}}<0.3706$
which violates the classical inequality $\alpha \geq 1$  \cite{Grangier86}.
The limits are given by data for $n_{\mathrm{rel}}=1$ in Table II.
All the above parameters are monotonously decreasing as less noise is imposed by LD1 and LD2
and for $n_{\mathrm{rel}}=0.2$ we already have $\alpha=0.0521$ and  $g^{(2)}(0)=0.0519$.

In conclusion, we have examined a source producing approximate single-photon states with positive Wigner function but exhibiting
strong photon anti-correlation and we have unambiguously proved that the generated states 
cannot be expressed as mixtures of Gaussian states.
In comparison  to the witness based on negativity of the Wigner function \cite{Silberhorn}, 
the present criterion can identify a high nonclassicality of a much wider class of single photon sources, 
while avoiding the need for demanding estimation of complete photon number distribution or complicated data processing \cite{Mari11}. 
Consequently, the presented criterion is particularly useful for evaluation of single-photon sources 
where negativity of Wigner function cannot be observed \cite{Scheel}.

\begin{acknowledgments}
The work was supported by Projects No. MSM6198959213, No. LC06007 and ME10156
of the Czech Ministry of Education, by Palacky University (PrF-2011-015) and by Czech Science Foundation (202/09/0747).
\end{acknowledgments}


\begin{thebibliography}{99}
\bibitem{note1}
L. Lachman and R. Filip, in preparation; using the same structure of the witness, the non-classicality witness 
can be expressed as $W(a)>e^{a-1}/(1+a)$.
\end{thebibliography}
\end{document}